# Explanation and exact formula of Zipf's law evaluated from rank-share combinatorics

A Shyklo

**ABSTRACT**
This work proves that ranks and shares are statistically dependent on one another, based on simple combinatorics. It presents a formula for rank-share distribution and illustrates that Zipf's law, is descended from expected values of various ranks in the new distribution. All conclusions, formulas and charts presented here were tested against publically available statistical data in different areas. The correlation coefficient between the calculated values and statistical numbers provided by Bureau of Labor Statistics was 0.99899. Monte-Carlo simulations were performed as additional evidence.

**Introduction**
The mysterious Zipf's law astonishes researchers for over 100 years already. It was initially presented by Jean-Baptiste Estoup [1] in 1908. He observed a strange proportional dependency between frequencies of word usage in texts. Later it was observed in many languages, that the frequency of most common words is proportional to 1/rank. For example, the word "the" is the most commonly used word in the English language. The second most common, "of" is used about half as much as the first. The third, "and" is used about a third as much as the first, and so on.

This dependency was popularized by and named after a linguist from Harvard University known as George Kingsley Zipf [2]. It was used in 1913 by German physicist Felix Auerbach in the "Law of Population Concentration" to describe the size distribution of cities. In 1991, Wentian Li demonstrates [3] that randomly generated texts follow the same frequency distribution as real languages.

The pattern is distinct in a great deal of research, which causes a recognizable empirical distribution. It can be found in the use of words, in city populations, last names, distribution of wealth, frequency of natural disasters, markets behavior etc. It is distinct in the 80/20 rule.

There were multiple attempts to explain it. And it was partially done in many publications [4], [5], [6], [7], [8], [9], [10], [11], [12], [13], [14], [15], [16], [17], however the exact math behind it remained unknown, even after centuries of research.

This work started as a practical attempt to apply the latest statistical formulas to real life data. Working with large datasets, we surprisingly found inaccuracy in the existing equations. Closer observations of the various data samples revealed statistical dependency between rank, share and number of participants. Further analysis led to a solid understanding of the combinatorics driving ranking process and exact formula for rank-share distribution, which provides key to understanding of Zipf's law and Pareto principle.

**Results**
To demonstrate the dependency between the rank and share, let's assume that we have combined volume T shared between N participants. Using combinatorics principles we can calculate that there are

$$\frac{(T+N-1)!}{(N-1)!T!}$$ ways to split the volume.

If we sort and rank each case and count how many times the share of some rank equal to a certain number (S), we can calculate the probability of this event. If the outcome appears x times, (when the rank k has the share S), the probability of this outcome can be calculated as:

$$P(T,N,k,S) = x * \frac{(N-1)!T!}{(T+N-1)!}$$

To illustrate it let's look at this simplified example:
Let's assume that we had 3 companies, which sold 10 items combined. There are 66 possible combinations of how they can split the market volume (T =10). If we sort each combination and count how many times rank one has each value from 0 to 10, we can create the following chart:

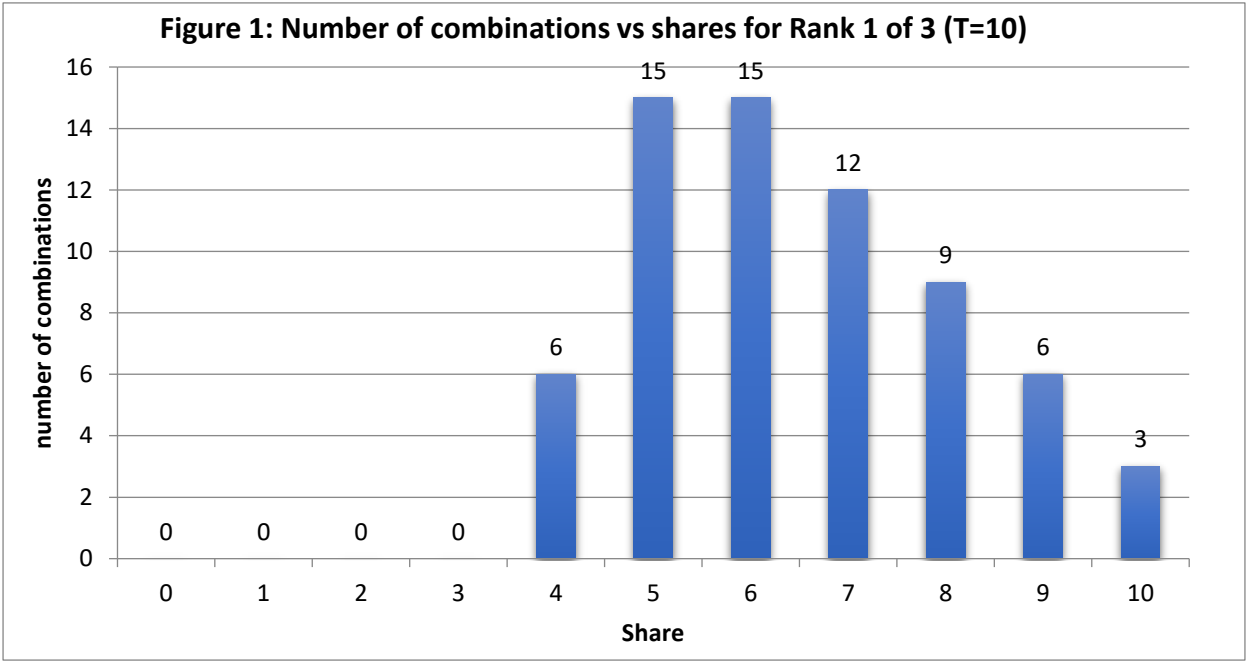

From this chart we can see, for example, that the probability of the Rank 1 to have the share = 7 of 10 (or 70%) is 12/66 or 18.1818%. So we can calculate the probability of every share value for rank 1
Similar charts can be created for Rank2 and Rank3.

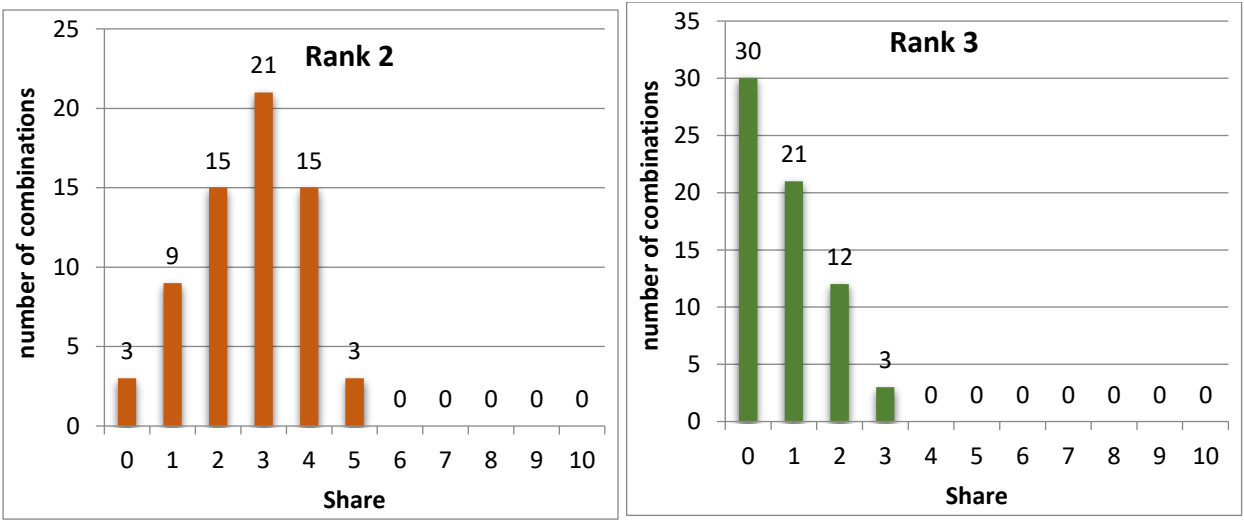

Using this logic we can calculate Probability Density Functions of various ranks, N and T.

**Figure 3: Probability Density Functions for all ranks (N from 2 to 5)**

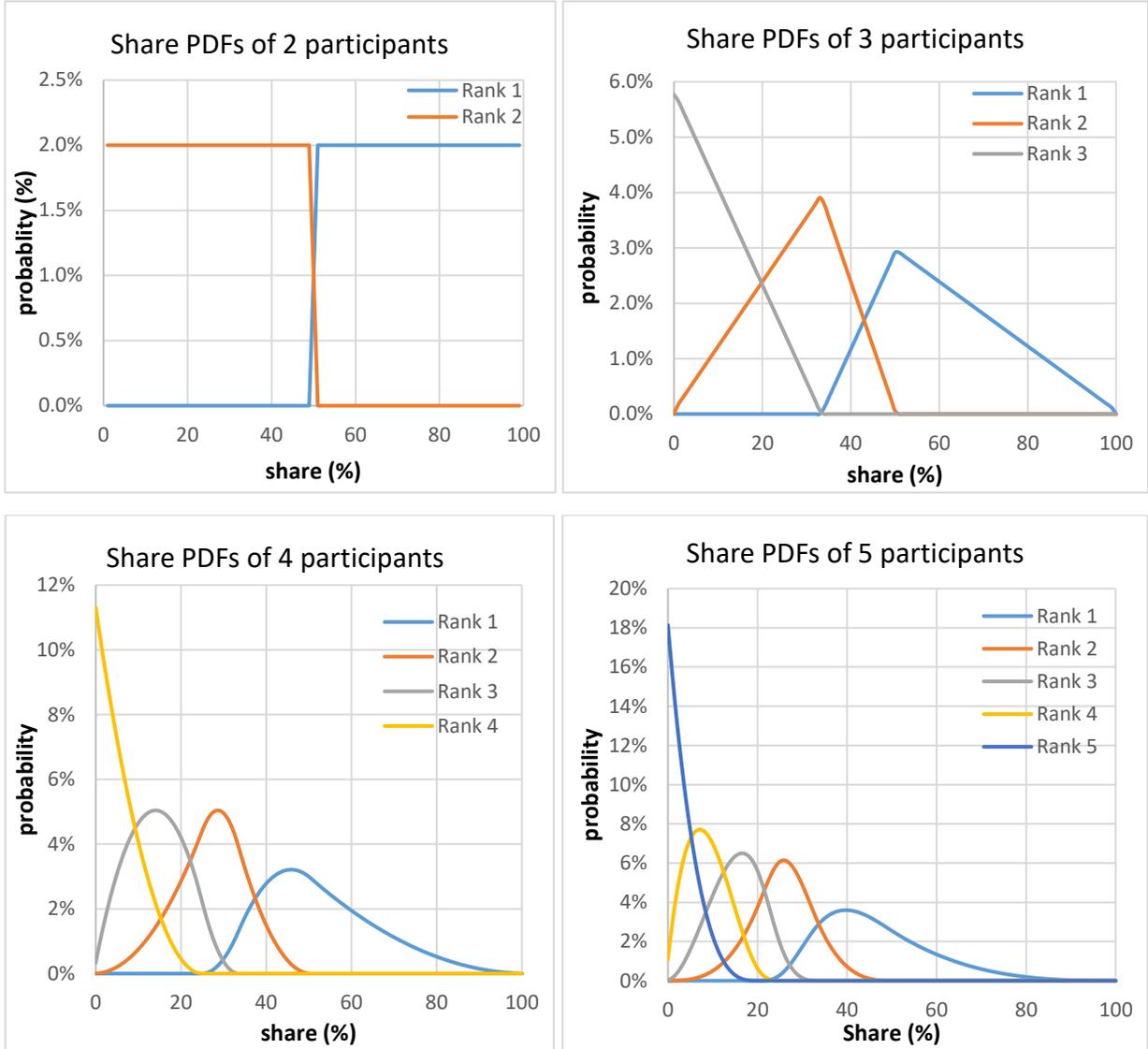

We were able to evaluate the universal formula for PDFs of rank-share distribution as:

$$P(N,k,S) = \frac{(N-1)!N!}{(k-1)!(N-k)!} * \sum_{i=k}^{\min\{N,\ \text{floor}(1/S)\}} \left( (-1)^{i-k} * \frac{(N-k)!}{(i-k)!(N-i)!} * (1-iS)^{N-2} \right)$$

To prove this formula, we tested it against statistical data and performed Monte-Carlo simulations. Figure 4 represents the same distributions for N = 4 calculated based on 500000 random outcomes.

**Figure 4: Monte-Carlo Simulation of N = 4**

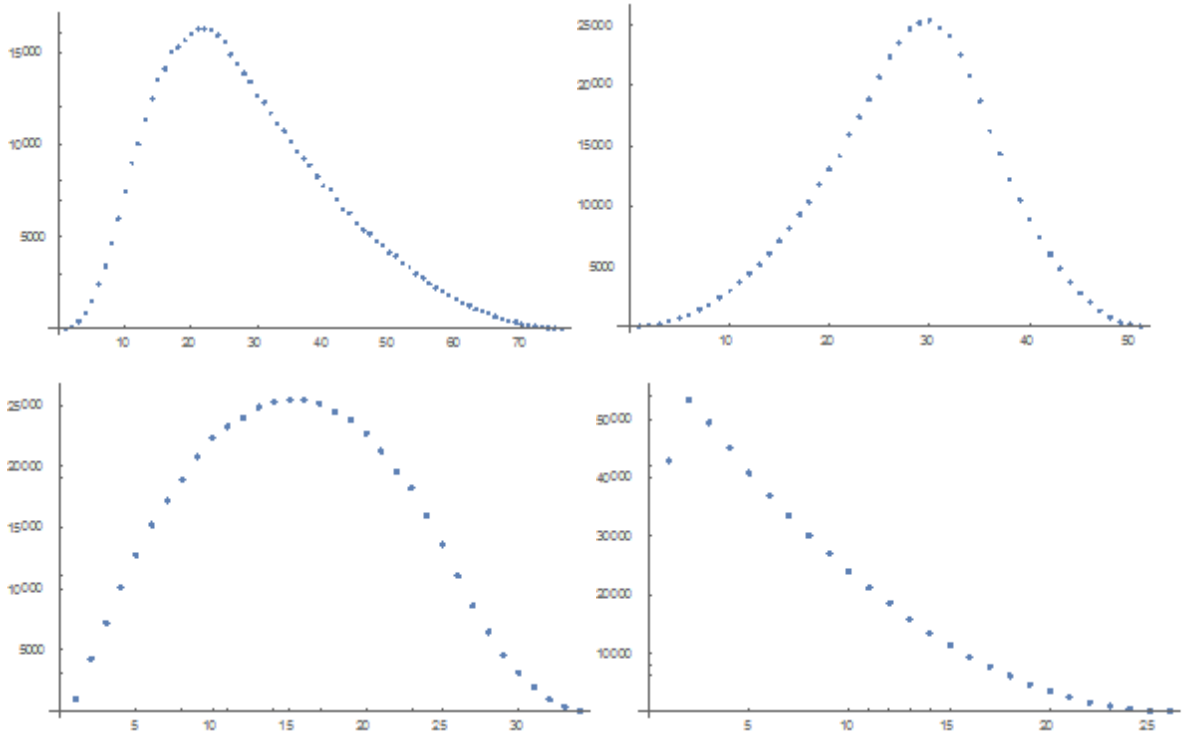

The expected values of share for each rank can be calculated as:

$$S = \frac{1}{N} * \left(\sum_{i=k}^{N} \frac{1}{i}\right)$$

Where S represents share for rank k, and N - number of participants

**Figure 5: Expected values of the rank-share distribution.**

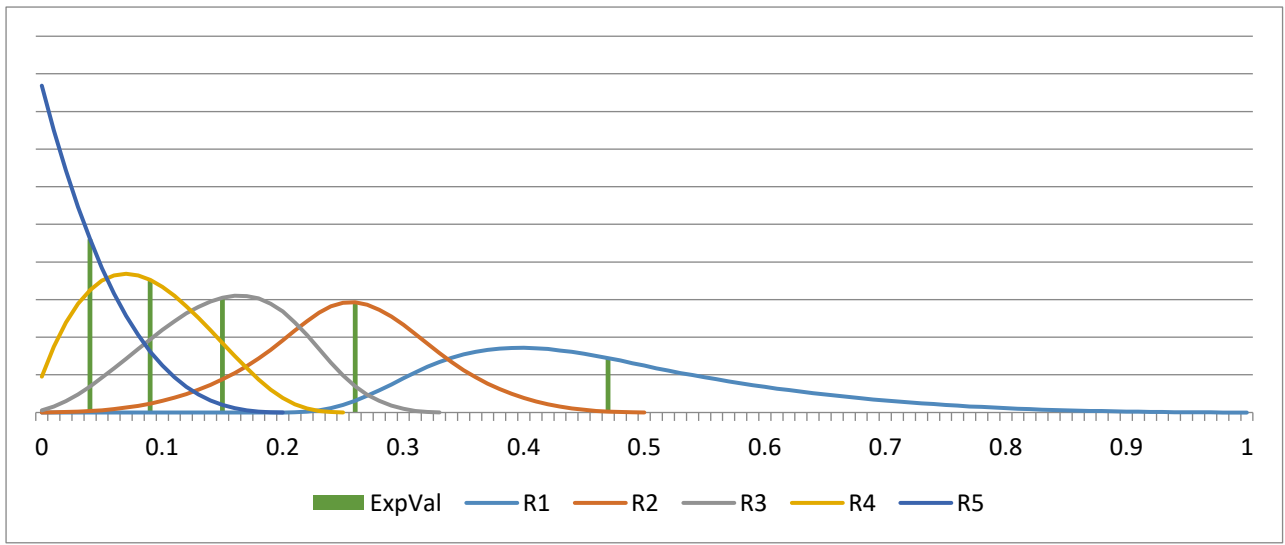

**Figure 6: Expected values for Zipf's law in double log scale calculated for N from 1 to 100**

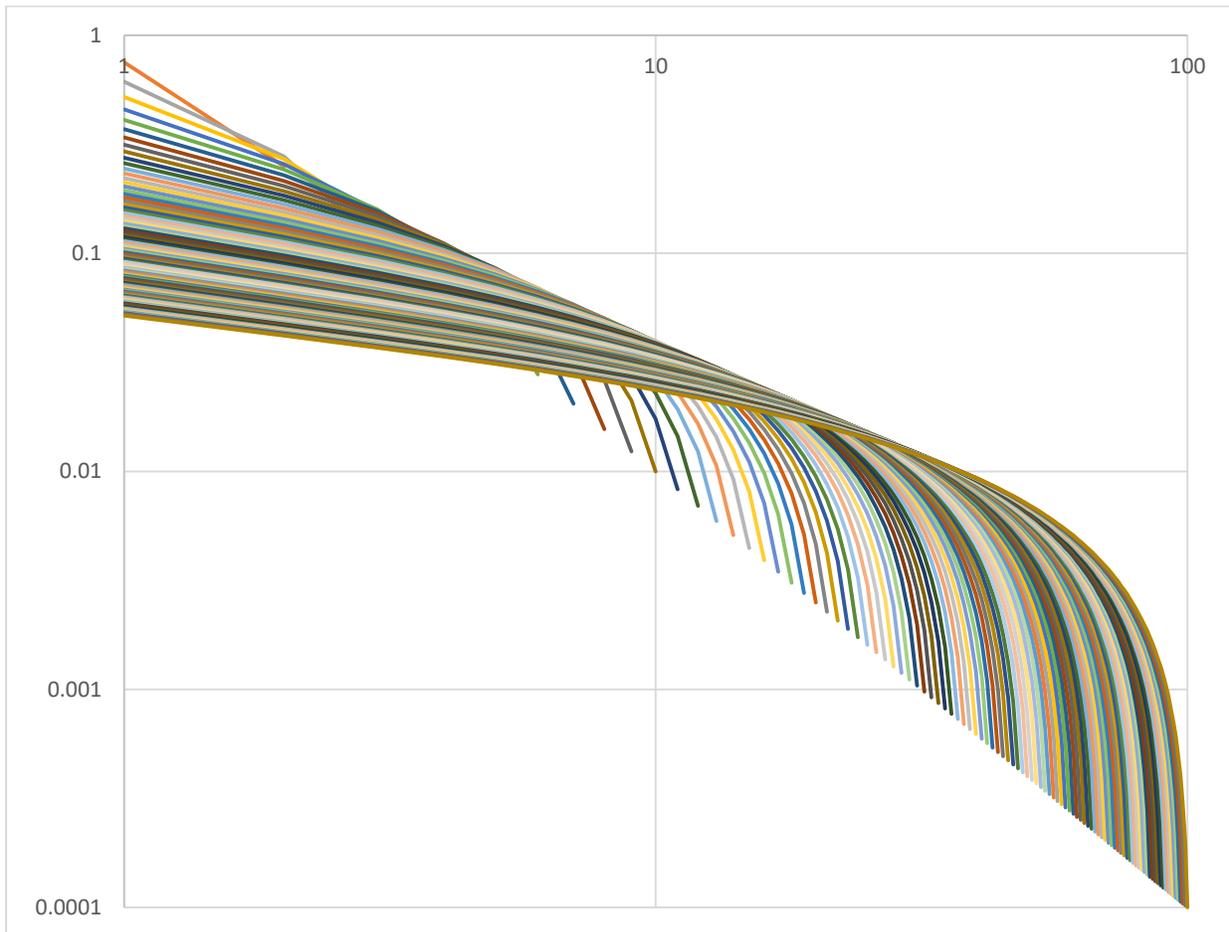

The expected values of the rank-share distribution gives us the dependency between rank and frequency known as Zipf's Law. To prove it we tested it on publically available datasets with known N. For example we know that Canada has 13 states, Brazil has 27 states and US – 50 States. We can get statistics of the area distribution for each country from these sources [18],[19],[20].

**Figure 7: Shares of the area of the states in US, Canada, Brazil on double log scale (real vs calc.)**

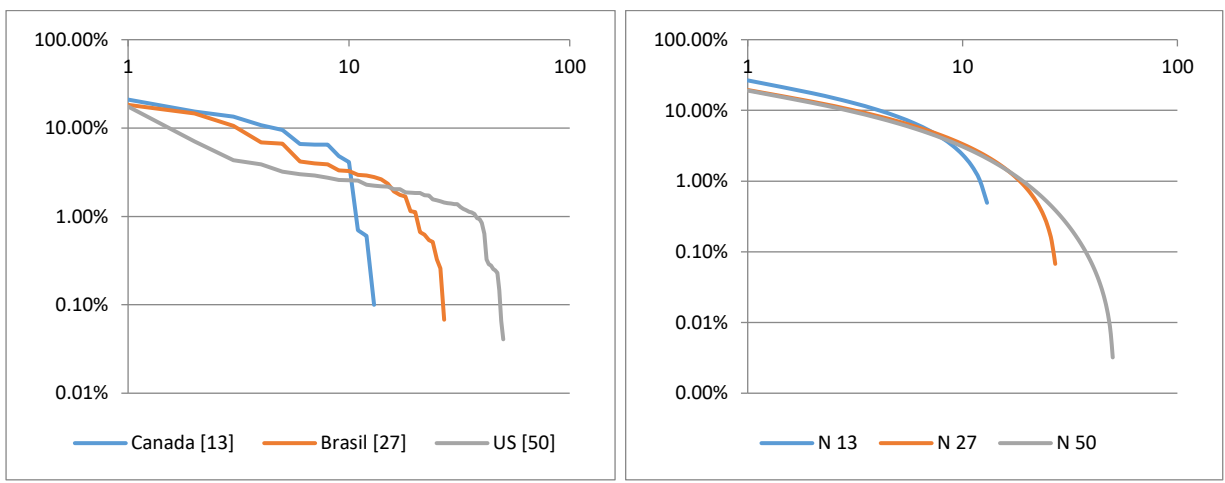

Another example can be the distribution of letters among European languages, published here [21]. We exactly know how many letters are in each language.

**Figure 8: Frequencies of letters usage in languages on double log scale (real vs expected values).**

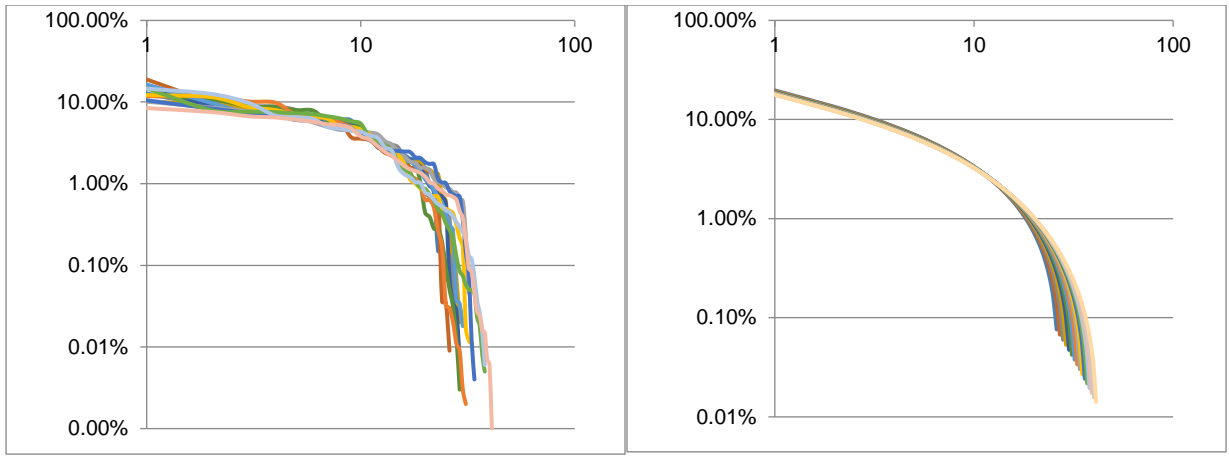

**Table 1: Correlation coefficients between real and calculated distributions of letters usage.**

| Language | Correlation | Language | Correlation |
| --- | --- | --- | --- |
| Czech | 0.972696909 | Icelandic | 0.984039287 |
| Danish | 0.983906839 | Italian | 0.974405202 |
| Dutch | 0.972386224 | Polish | 0.99029575 |
| Esperanto | 0.979436537 | Portuguese | 0.984913919 |
| Finnish | 0.977592954 | Spanish | 0.985479455 |
| French | 0.971366449 | Swedish | 0.969368024 |
| German | 0.98786782 | Turkish | 0.991741606 |

To achieve a more accurate verification, we bundled together numbers provided by Bureau of Labor Statistics of US Department of Labor [22]. We analyzed Occupational Employment and Wages distributions between 22 categories in more than 50 US cities. We ranked the categories for each city, then we calculated the average share for each rank from 1 to 22, and compared results to the calculated values. The observed correlation coefficient was 0.99899712.

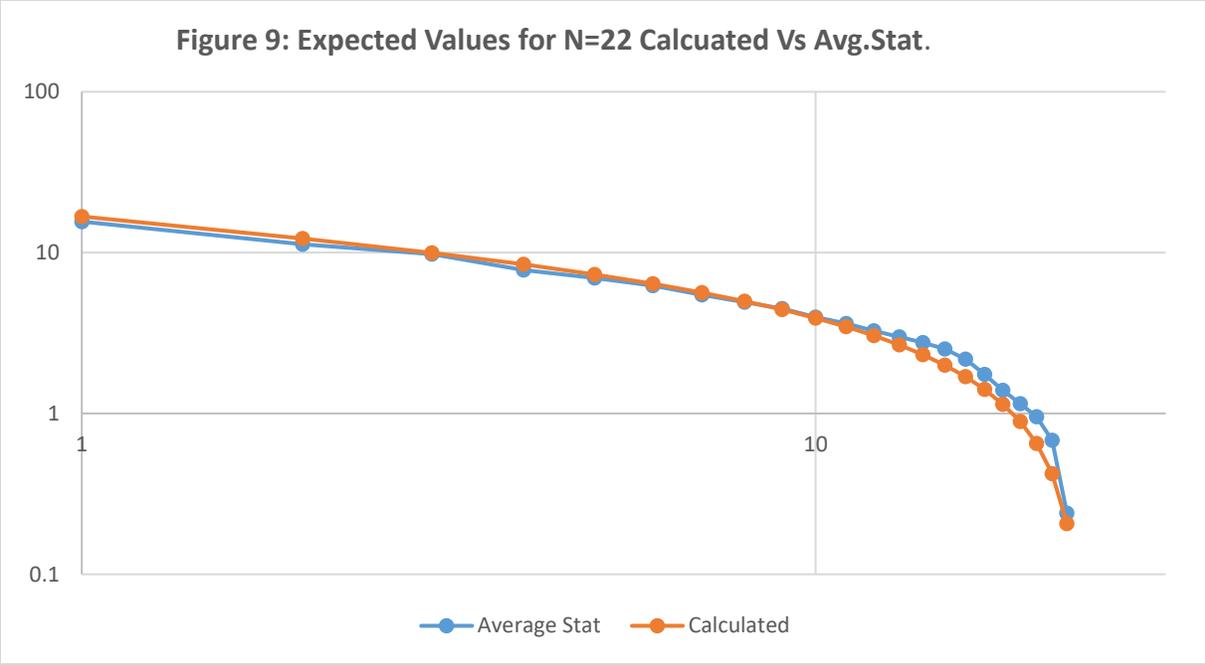

Figure 9: Expected Values for N=22 Calcuated Vs Avg.Stat.

## Conclusion
We demonstrated that rank and share are statistically related and evaluated an exact formula for the rank-share distribution. This is a universal law, which can be applied to any area. That's why we can find it in completely different places (words, population, markets). The expected values of the new distribution gives us the dependency between rank and frequency known as Zipf's Law. We can rank distribution of words, frequency of natural disasters, population in the cities or income spreading, and observe the same pattern caused by simple rank-share combinatorics.

## Discussion
The PDF formula of the rank-share distribution contains binomial coefficients and probably could be simplified using binomial equations. The rank-share distribution possibly belongs to binomial series. It still needs to be classified. We spend significant time trying to derive it from known distributions. We were able partially derive dependency for some ranks between rank-share and Negative-Binomial distribution. However, universal dependency still should be evaluated.   .

For this work we were concentrating on continuous solutions, assuming that T (number of shared items) is large enough to be considered as ∞, but it would be interesting to derive an exact formula for discrete solutions including T as a parameter.

## Methods
We started analyzing big sets of statistical data and observed strong recognizable patterns between ranks, shares and number of participants. We also noticed that possible share values for each rank were located within certain range and figured the logic for ranges:

> The maximum share of Rank k is always 1/k.
> The minimum share is 0 for all ranks except Rank 1 (where min is 1/N).

To explain the logic behind it, imagine the case when we have just two participants. They split shares in some proportions (50/50, 60/40 or so.). Participant ranked #1 could not have a share less than 50%. The same way participant #2 could not have a share more than 50%. So #1 distributed between 1/2 and 1 while #2 between 0 and 1/2. In the case of 3 players, the lowest possible value for #1 is 1/3 (case of equal distribution between #1, #2 and #3). The highest possible value for #2 is 1/2 (case50/50/0). As a result we have #1 between 1/3 and 1. #2 between 0 and 1/2. #3 between 0 and 1/3.

We tested this logic for many N and it perfectly fit with real data.

At some point we realized that combinatorics may also govern distributions of shares for each rank. To test it we created a simple python algorithm. We used N nested loops with N variables and counted only cases where sum of all variables equal T. Then we could rank each case indexing all the shares used for each rank.
We must acknowledge that there can be variations among the methodologies of the ranking process. For example: What should we do in cases when the shares of two participants are equal? Should we consider participants with the share = 0? We did a relatively complicated impact analysis of these factors and observed that when T is big enough, all scenarios will lead us to the same dependency.

The outcomes of our calculation correlates with real statistical data.
We continued seriously testing it. Tests were conducted for various ranks up to N=70.
To achieve it we completed significant work improving the effectiveness of algorithms, using various programming environments and came up with a very effective recursive C# algorithm scalable for map reduce, which let us process the cases up to N= 100 and T = 5000.
We combined all pre-calculated cases in the database and tried to fit the data to one of the existing PDFs. Unfortunately, we were not able to find proper distribution or to derive our calculated numbers from known distributions. Thus we started working on a universal formula for rank-share PDF. We tested our approaches against a pre-calculated database.

### Continuous Formula for the Last Rank

Assume we have 5 participants (N = 5). How can we count the number of combinations for Rank 5?

For a given value of $S_5$, the minimum value of $S_4$ can be $S_5$, the maximum is $(T-S_5)/(N-1)$.
For each $S_4$, the minimum value of $S_3$ is $S_4$, the maximum is $(T-S_5-S_4)/(N-2)$
For each $S_3$, the minimum value of $S_2$ is $S_3$, the maximum is $(T-S_5-S_4-S_3)/(N-3)$.
For each $S_2$, we have just one value of $S_1 = T-S_2-S_3-S_4-S_5$.

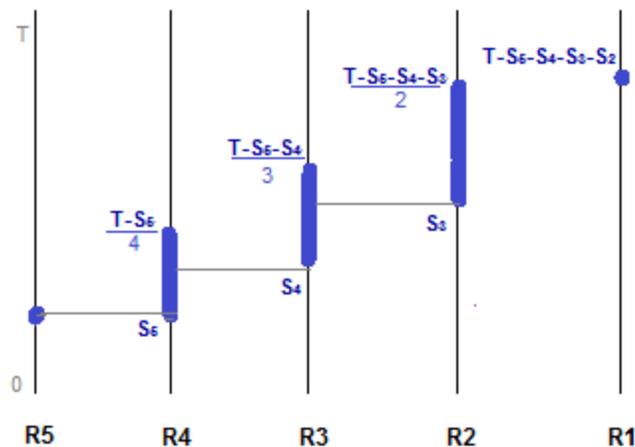

Figure 10: Combinations for last rank of 5

Based on this logic we can calculate the quantity (number of combination) for the Last Rank as:

$$Q(S_N) = \int_{S_N}^{\frac{(T-S_N)}{(N-1)}} \int_{S_{N-1}}^{\frac{(T-S_N-S_{N-1})}{(N-2)}} \cdots \iiint \cdots \int_{S_2}^{\frac{(T-\sum_{i=2}^{N}S_i)}{1}} 1 \; dS_1 \; dS_2 \ldots dS_{N-2} \; dS_{N-1}$$

We used SymPy Python package to calculate results of these interactions and found the pattern:

$$Q(T, N, S_N) = \frac{(T - NS_N)^{(N-2)}}{(N-2)!(N-3)!}$$

When we normalize it we can calculate the probability of the last rank

$$P(T, N, S_N) = \frac{(N-1)N}{T^N} * (T - NS_N)^{(N-2)}$$

The expected value of the last rank can be calculated as $\frac{1}{N^2}$

### Continuous Formula for N-1 Rank

In our example, for the second lowest rank (N-1), we should start from $S_4$. The right part of the diagram remains the same. In the left part, the minimum of $S_5$ is 0, but the maximum can be either $S_4$ or $(T-S_4*4)$, depending on what is less.

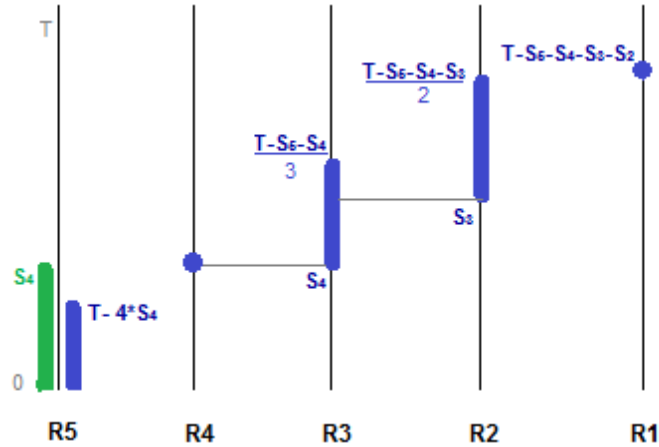

**Figure 11: Combinations for rank 4 of 5**

There are two different equations depending if $S_4 > (T-S_4*4)$. We can calculate the results for both integral equations. For continuous solutions, we can represent the quantities for $S_4$ of N=5 as two polynomials:

$$-\frac{64}{3} * S_4^3 + (4S_4)^2 - 4S_4 + \frac{1}{3}$$

$$\frac{(1-4S_4)^3}{3}$$

We can also analyze the condition and see that the first polynomial works for S₄ < 1/5 and the second one for S₄ >= 1/5.

For this example, the continuous solution can be presented as following graph:

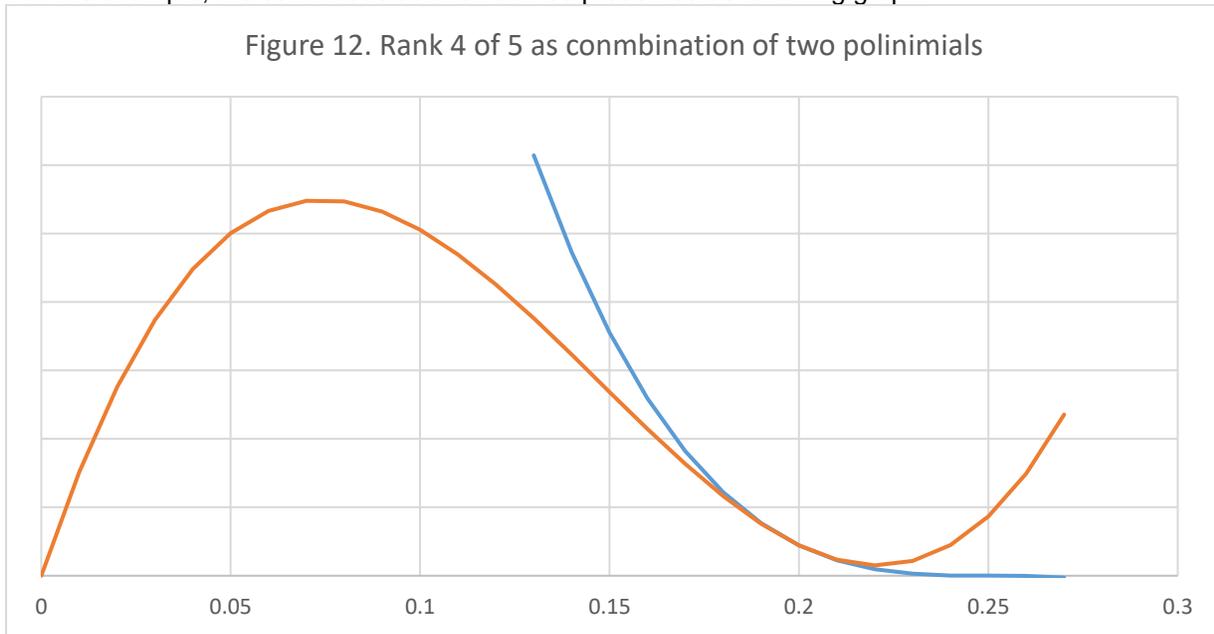

Figure 12. Rank 4 of 5 as conmbination of two polinimials

The universal formula for the first polynomial, applied on the interval between 0 and 1/N, is:

$$\frac{(1-(N-1)*S_{N-1})^{(N-2)} - (1-N*S_{N-1})^{(N-2)}}{N-2}$$

The universal formula for the second polynomial, applied on the interval between 1/N and 1/(N-1), is:

$$\frac{(1-(N-1)*S_{N-1})^{(N-2)}}{N-2}$$

To get exact PDF formulas, we should also normalize the equations.
We can also calculate the expected values for rank N-1 as $\frac{1}{N(N-1)}$

### Middle Ranks Formulas

We can follow this path to see that the higher the rank, the more integrals we need for the continuous solution. The solution of these integrals would be a set of polynomials. There are different functions on the intervals 1 to 1/2, 1/2 to 1/3, 1/3 to 1/4, 1/4 to 1/5 …. In general, the PDF for N-tier ranks can be represented as sets of polynomial functions with the degree (N-2).
For example assuming T=1, we can calculate the not normalized polynomials for N =3 to 5 as:

**Table 2: Polynomials for share PDS's calculated for N from 3 to 5**

| S (for N=3) | R1 | R2 | R3 |
|---|---|---|---|
| 0 - 1/3 |  | 2S | -(3S-1) |
| 1/3-1/2 | (3S-1) | -2(2S-1) |  |
| 1/2 - 1 | (1-S) |  |  |

| S (for N=4) | R1 | R2 | R3 | R4 |
|---|---|---|---|---|
| 0 - 1/4 |  | $6S^2$ | -3S(7S-2) | $(4S-1)^2$ |
| 1/4-1/3 | $(4S-1)^2$ | $-42S^2+24*S-3$ | $3(3S-1)^2$ |  |
| 1/3-1/2 | $-11S^2+10S-2$ | $3(2S-1)^2$ |  |  |
| 1/2 - 1 | $(1-S)^2$ |  |  |  |

| S (for N=5) | R1 | R2 | R3 | R4 | R5 |
|---|---|---|---|---|---|
| 0 - 1/5 |  | $24S^3$ | $36S^2*(1-4S)$ | $4S(61S^2-27S+3)$ | $(1-5S)^3$ |
| 1/5-1/4 | $(5S-1)^3$ | $-476S^3+300S^2-60S+4$ | $4((1-3S)^3 - 2(1-4S)^3)$ | $4(1-4S)^3$ |  |
| 1/4-1/3 | $-131S^3 + 117S^2 - 33S + 3$ | $(1-2S)^3 - 3(1-3S)^3$ | $6(1-3S)^3$ |  |  |
| 1/3-1/2 | $(1-S)^3 - 4(1-2S)^3$ | $4(1-2S)^3$ |  |  |  |
| 1/2 - 1 | $(1-S)^3$ |  |  |  |  |

Here are some more graphical representations:

**Figure 12: Graphical representations of polynomials for PDF's for ranks 2 and 3 of 5**

**N5 R3 polynomials**

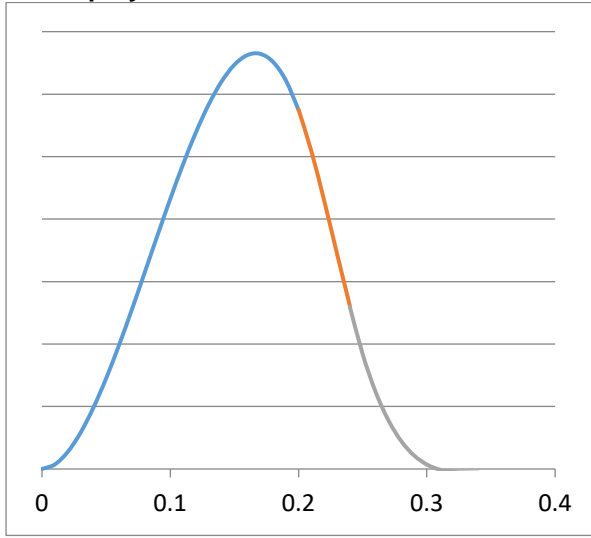

**N5 R2 polynomials**

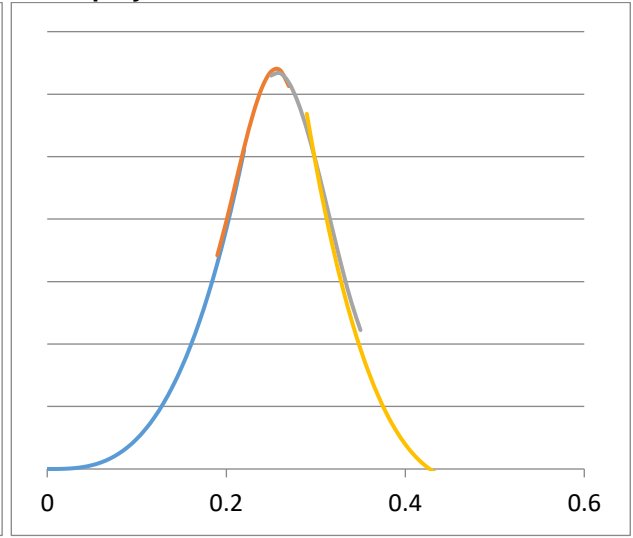

**Universal formula for PDF.**

When we analyzed polynomials for ranks N and N-1 we realized that they could be presented as a sum of the terms like $(1-iS)^{(N-2)}$ multiplied by coefficients ($a_1$ $a_2$ $a_3$ $a_4$ $a_5$), where $i$ between 1 and N.

$$P(N,k,S) \sim [a_1 a_2 a_3 a_4 a_5 ...] * \begin{bmatrix} (1-1S)^{(N-2)} \\ (1-2S)^{(N-2)} \\ (1-3S)^{(N-2)} \\ (1-4S)^{(N-2)} \\ (1-5S)^{(N-2)} \\ ... \end{bmatrix}$$

To calculate coefficients for all polynomial equations we created the Python package, which parsed through all combinations of coefficients and returned a corresponding matrix for the given polynomial. These are the results calculated for all polynomials with N from 3 to 5:

**Table 3: Coefficients for polynomial equations calculated for N from 3 to 5**

| S (for N=3) | R1 | R2 | R3 |
|---|---|---|---|
| 0 - 1/3 |  | [0 2 -2] | [0 0 1] |
| 1/3-1/2 | [1 2 0] | [0 2 0] |  |
| 1/2 - 1 | [1 0 0] |  |  |

| S (for N=4) | R1 | R2 | R3 | R4 |
|---|---|---|---|---|
| 0 - 1/4 |  | [0 3 -6 3] | [0 0 3 -3] | [0 0 0 1] |
| 1/4-1/3 | [1 -3 3 0] | [0 3 -6 0] | [0 0 3 0] |  |
| 1/3-1/2 | [1 -3 0 0] | [0 3 0 0] |  |  |
| 1/2 - 1 | [1 0 0 0] |  |  |  |

| S (for N=5) | R1 | R2 | R3 | R4 | R5 |
|---|---|---|---|---|---|
| 0 - 1/5 |  | [0 4 -12 12 4] | [0 0 6 -12 6] | [0 0 0 4 -4] | [0 0 0 0 1] |
| 1/5-1/4 | [1 -4 6 -4 0] | [0 -4 12 -12 0] | [0 0 6 -12 0] | [0 0 0 4 0] |  |
| 1/4-1/3 | [1 -4 6 0 0] | [0 4 -12 0 0] | [0 0 6 0 0] |  |  |
| 1/3-1/2 | [1 -4 0 0 0] | [0 4 0 0 0] |  |  |  |
| 1/2 - 1 | [1 0 0 0 0] |  |  |  |  |

It's obvious that the coefficients follow a binomial pattern. When we normalize the dependency, we evaluate the following formula for PDFs of rank-share distribution:

$$P(N,k,d,S) = \frac{(N-1)!N!}{(k-1)!(N-k)!} * \sum_{i=k}^{d} \left( (-1)^{i-k} * \frac{(N-k)!}{(i-k)!(N-i)!} * (1-iS)^{N-2} \right)$$

Where S represents the share for rank k, N numbers of participants,
d represents range  (like d=1 [1/2 – 1]
            d=2 [1/3 – 1/2]
            d=3 [1/4 – 1/3] …N)
d is related to S and could not be more than N. So it can be calculated as min$\{N, \lfloor 1/S \rfloor\}$

### Verification

To verify the equations we tested them against publically available datasets from various sources [18],[19],[20],[21],[22] with known number of categories. We ranked and normalized each dataset to fit shares between 0 and 100%. For example, we used data extracts from Bureau of Labor Statistics of US Department of Labor.

**Table 4: Example of occupational employment and wages dataset for various US towns.**

| Major occupational group | Percent of total employment | | | | |
|---|---|---|---|---|---|
| | Birmingham | Montgomery | Anchorage | Fairbanks | Flagstaff |
| **Total, all occupations** | 100.00% | 100.00% | 100.00% | 100.00% | 100.00% |
| Management | 4.2* | 3.8* | 5.7* | 5.9* | 5.5 |
| Business and financial operations | 4.4* | 4.5* | 5.1 | 3.6* | 3.3* |
| Computer and mathematical | 2.6* | 2.4* | 1.9* | 1.7* | 1.3* |
| Architecture and engineering | 1.4* | 1.6* | 3.0* | 2.2 | 1.4 |
| Life, physical, and social science | 0.5* | 0.7* | 1.5* | 3.1* | 2.7* |
| Community and social services | 0.8* | 1.2* | 2.0* | 1.7* | 1.8* |
| Legal | 0.8 | 0.8 | 0.8 | 0.6* | 0.7 |
| Education, training, and library | 5.2* | 6 | 5.2* | 8.1* | 6.9 |
| Arts, design, entertainment, sports, and media | 1.1* | 1.1* | 1.3 | 0.9* | 1.2 |
| Healthcare practitioner and technical | 7.9* | 5.3 | 5.7 | 5.7 | 6.3* |
| Healthcare support | 2.7 | 2.4* | 2.9 | 2.1* | 1.8* |
| Protective service | 2.7* | 3.1* | 2.3* | 2.1 | 2.9* |
| Food preparation and serving related | 8.1* | 8.8 | 9.3 | 8.9 | 15.0* |
| Building and grounds cleaning and maintenance | 2.6* | 3.5 | 3 | 4.3 | 4.2* |
| Personal care and service | 2.6* | 2.8* | 4.5* | 1.8* | 4.0* |
| Sales and related | 12.5* | 10.6 | 9.3* | 8.3* | 10.7 |
| Office and administrative support | 16.9* | 15.6 | 17.2* | 16.3 | 14.1* |
| Farming, fishing, and forestry | 0.1* | 0.4 | 0.1* | 0.2* | 0.1* |
| Construction and extraction | 4.1 | 3.0* | 5.5* | 8.1* | 3.2* |
| Installation, maintenance, and repair | 4.7* | 4 | 4.6* | 5.5* | 4.7* |
| Production | 6.6 | 10.4* | 2.1* | 2.3* | 4.0* |
| Transportation and material moving | 7.3 | 7.9* | 7.1 | 6.6 | 4.4* |

The data was be transformed to calculate expected values for all rank from 1 to 22 like this:

**Table 5: Example of transformed and ranked data to evaluate expected values for shares (N=22)**

| Rank | Birmingham | Montgomery | Anchorage | Fairbanks | Flagstaff | Exp. Value. |
|---|---|---|---|---|---|---|
| 1 | 16.9 | 15.6 | 17.2 | 16.3 | 15 | **16.2** |
| 2 | 12.5 | 10.6 | 9.3 | 8.9 | 14.1 | **11.08** |
| 3 | 8.1 | 10.4 | 9.3 | 8.3 | 10.7 | **9.36** |
| 4 | 7.9 | 8.8 | 7.1 | 8.1 | 6.9 | **7.76** |
| 5 | 7.3 | 7.9 | 5.7 | 8.1 | 6.3 | **7.06** |
| 6 | 6.6 | 6 | 5.7 | 6.6 | 5.5 | **6.08** |
| 7 | 5.2 | 5.3 | 5.5 | 5.9 | 4.7 | **5.32** |
| 8 | 4.7 | 4.5 | 5.2 | 5.7 | 4.4 | **4.9** |
| 9 | 4.4 | 4 | 5.1 | 5.5 | 4.2 | **4.64** |
| 10 | 4.2 | 3.8 | 4.6 | 4.3 | 4 | **4.18** |
| 11 | 4.1 | 3.5 | 4.5 | 3.6 | 4 | **3.94** |
| 12 | 2.7 | 3.1 | 3 | 3.1 | 3.3 | **3.04** |
| 13 | 2.7 | 3 | 3 | 2.3 | 3.2 | **2.84** |
| 14 | 2.6 | 2.8 | 2.9 | 2.2 | 2.9 | **2.68** |
| 15 | 2.6 | 2.4 | 2.3 | 2.1 | 2.7 | **2.42** |
| 16 | 2.6 | 2.4 | 2.1 | 2.1 | 1.8 | **2.2** |
| 17 | 1.4 | 1.6 | 2 | 1.8 | 1.8 | **1.72** |
| 18 | 1.1 | 1.2 | 1.9 | 1.7 | 1.4 | **1.46** |
| 19 | 0.8 | 1.1 | 1.5 | 1.7 | 1.3 | **1.28** |
| 20 | 0.8 | 0.8 | 1.3 | 0.9 | 1.2 | **1** |
| 21 | 0.5 | 0.7 | 0.8 | 0.6 | 0.7 | **0.66** |
| 22 | 0.1 | 0.4 | 0.1 | 0.2 | 0.1 | **0.18** |

For verification, we combined data from more than 50 US towns.

**Monte-Carlo Simulation.**

We used Wolfram Mathematica to perform Monte Carlo simulations for PDFs (for N from 2 to 6).
The following code was used:

```
m = RandomInteger[100, {500000, 3}]
m2 = Sort /@ m
m3 = Transpose[{m2[[All, 1]], m2[[All, 2]] - m2[[All, 1]],  m2[[All, 3]] - m2[[All, 2]], 100 - m2[[All, 3]]}]
m4 = Sort /@ m3
ListPlot[Values[KeySort[Counts[m4[[All, 4]]]]]]
```

**Figure 13: Monte-Carlo Simulation for N3 and N5**

**N3:**

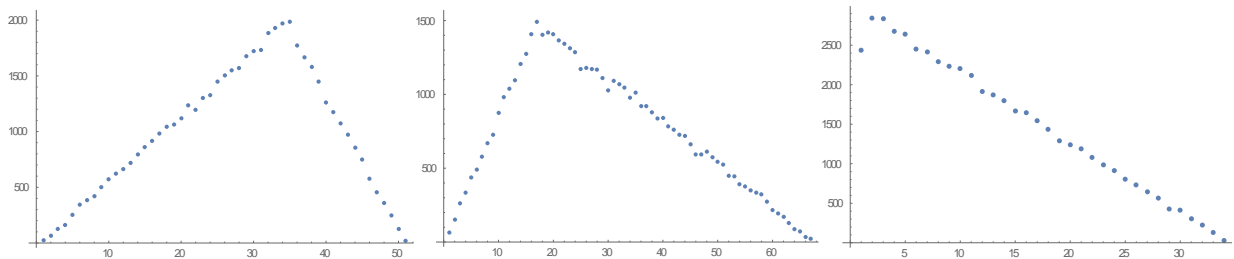

**N5:**

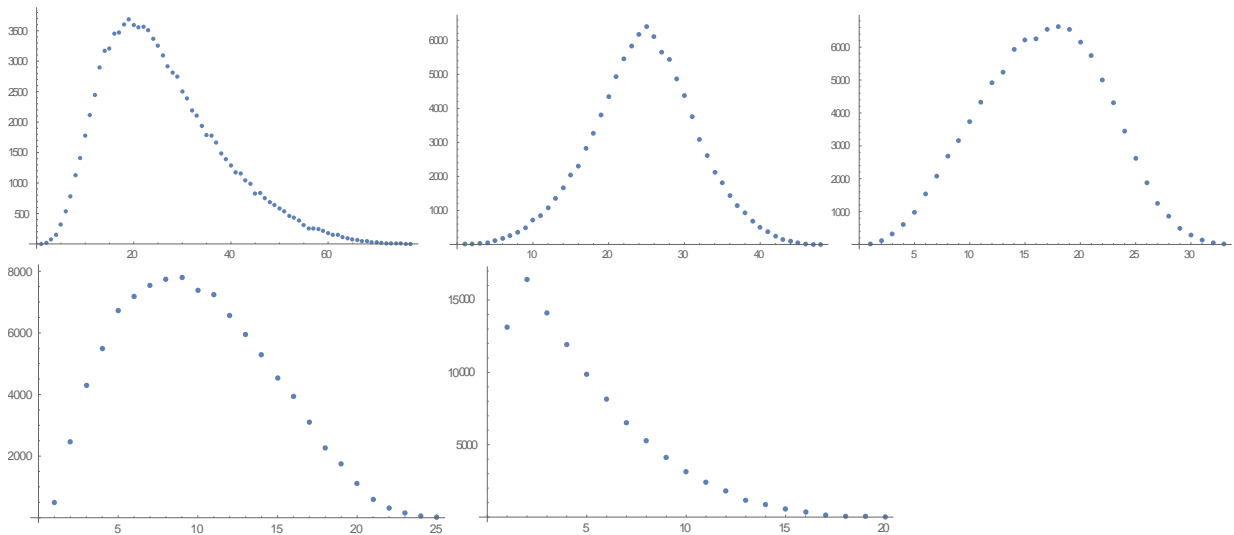

## References

1. J. Estoup. Gammes sténographiques. Institut Stenographique de France, 1916.

2. Zipf, G.K. Selected Studies of the Principle of Relative Frequency in Language. Cambridge, MA: Harvard University Press. ISBN 9780674434929 (1932).
http://www.hup.harvard.edu/catalog.php?isbn=9780674434929

3. W Li, Random texts exhibit Zipf's-law-like word frequency distribution. IEEE Transactions on Information Theory (1992). https://www.santafe.edu/research/results/working-papers/random-texts-exhibit-zipfs-law-like-word-frequency

4. Seung Ki Baek, Sebastian Bernhardsson , Petter Minnhagen. Zipf 's law unzipped. arXiv.org. (2011)
https://arxiv.org/abs/1104.1789

5. Fernando Buendía. Market Shares Are Not Zipf-Distributed. Complex Systems. (2013)
http://www.complex-systems.com/pdf/22-3-2.pdf

6. Ruokuang Lin, Chunhua Bian, Qianli D.Y.Ma, Scaling laws in human speech, decreasing emergence of new words and a generalized model. arXiv.org. (2015). http://arxiv.org/pdf/1412.4846.pdf

7. Nikolay K. Vitanov, Marcel Ausloos, Test of two hypotheses explaining the size of populations in a system of cities. arXiv.org. (2015). http://arxiv.org/pdf/1506.08535.pdf

8. Vladimir V. Bochkarev, Eduard Yu. Lerner, Zipf and non-Zipf Laws for Homogeneous Markov Chain, . arXiv.org. (2012). http://arxiv.org/abs/1207.1872

9. Hila Riemer, Suman Mallik, Devanathan Sudharshan, Market Shares Follow the Zipf Distribution, College of Business at Illinois. (2002). http://www.business.illinois.edu/Working_Papers/papers/02-0125.pdf

10. Steven T. Piantadosi, Zipf's word frequency law in natural language: a critical review and future directions. The University of Rochester. (2015) https://colala.bcs.rochester.edu/papers/piantadosi2014zipfs.pdf

11. D.Yu. Namin, Mandelbrot's Model for Zipf's Law: Can Mandelbrot's Model Explain Zipf's Law for Language? Journal of Quantitative Linguistics 16(3):274-285 (2009)

12. AARON CLAUSET, COSMA ROHILLA SHALIZI, M. E. J. NEWMAN. POWER-LAW DISTRIBUTIONS IN EMPIRICAL DATA. arXiv.org. (2009) http://arxiv.org/pdf/0706.1062v2.pdf

13. Elvis Oltean. An econophysical approach of polynomial distribution applied to income and expenditure. arXiv.org. (2009) http://arxiv.org/ftp/arxiv/papers/1410/1410.3860.pdf

14. K. E. Kechedzhy, O. V. Usatenko, V. A. Yampol'skii, A. Ya. Usikov. Rank distributions of words in additive many-step Markov chains and the Zipf law. arXiv.org. (2004). http://arxiv.org/pdf/physics/0406099.pdf

15. Bernat Corominas-Murtra, Lus F Seoane, Ricard V. Sole. Zipf's law, unbounded complexity and open-ended evolution. arXiv.org. (2016) http://arxiv.org/abs/1612.01605

16. Oscar Fontanelli, Pedro Miramontes, Yaning Yang, Germinal Cocho, Wentian Li. Beyond Zipf's Law: The Lavalette Rank Function and Its Properties. arXiv.org. (2016). https://arxiv.org/pdf/1606.01959.pdf

17. Bohdan B. Khomtchouk, Claes Wahlestedt. Zipf's law emerges asymptotically during phase transitions in communicative systems. arXiv.org. (2016) https://arxiv.org/pdf/1603.03153.pdf

18. List of U.S. states and territories by area. Wikipedia. (2017). https://en.wikipedia.org/wiki/List_of_U.S._states_and_territories_by_area

19. List of Brazilian states by area. Wikipedia. (2016). https://en.wikipedia.org/wiki/List_of_Brazilian_states_by_area

20. List of Canadian provinces and territories by area. Wikipedia. (2017). https://en.wikipedia.org/wiki/List_of_Canadian_provinces_and_territories_by_area

21. Letter frequency. Wikipedia. (2017). https://en.wikipedia.org/wiki/Letter_frequency

22. Bureau of Labor Statistics of US Department of Labor. (2016) https://www.bls.gov/regions/
16